\documentclass[%
 aip,
 jcp,
 amsmath,amssymb,
 reprint
]{revtex4-2}
\usepackage{array}[=2016-10-06]
\usepackage{graphicx}
\usepackage{lipsum}
\usepackage{hyperref}
\usepackage{hhline}
\usepackage{array} 

\usepackage{xcolor}

\begin{abstract}
Predicting how nanoparticles interact in concentrated electrolytes remains a central challenge in nanoscience, with implications spanning energy storage, catalysis, and soft materials design. While Derjaguin–Landau–Verwey–Overbeek (DLVO) theory provides a widely used framework for modeling colloidal interactions, its electrostatic component relies on electric double-layer (EDL) theories whose validity in ionic liquids and concentrated electrolytes is consistently debated. Experiments have revealed significant deviations from classical screening behavior, including underscreening and oscillatory charge ordering, yet the consequences of these nonclassical effects for collective nanoparticle behavior remain unresolved.

Here, we demonstrate that the extended range and oscillatory nature of electrostatic screening in concentrated electrolytes predicts different phase behavior and morphology of colloidal ionogels, yielding distinct thermodynamic signatures, percolation thresholds, and amorphous network structures. Using molecular dynamics simulations with DLVO-type pair interactions parameterized by competing EDL models, we systematically compare the consequences of different screening theories on gel formation and configurations. Notably, the structural differences are largely invisible to conventional local descriptors but emerge clearly when quantified via topological data analysis.

These findings establish a direct link between unresolved nanoscale electrostatics and emergent colloidal organization, demonstrating that macroscopic observables, such as phase transitions and gel morphology, can serve as probes of interfacial ion ordering and electrostatic screening. More broadly, this work highlights the incompleteness of current EDL descriptions in concentrated electrolytes and provides a framework for using nanoparticle assembly to inform and refine electrostatic theories in complex ionic environments.
\end{abstract}

\begin{document}
\title{Colloidal gel morphology and phase behavior differs based on the electrostatic interaction model in concentrated electrolytes}
\author{Hwigwang Lim}
\affiliation{Department of Chemical and Biological Engineering, University of Wisconsin, Madison, WI, USA}
\author{Caleb J. Youngwerth}
\affiliation{Department of Physics, University of Wisconsin, Madison, WI, USA}
\author{Lisa Je}
\affiliation{Department of Chemical and Biological Engineering, University of Wisconsin, Madison, WI, USA}
\author{Matthew A.~Gebbie}
\affiliation{Department of Chemical and Biological Engineering, University of Wisconsin, Madison, WI, USA}
\author{Rose K. Cersonsky}
\email[]{rose.cersonsky@wisc.edu}
\affiliation{Department of Chemical and Biological Engineering, University of Wisconsin, Madison, WI, USA}

\maketitle

\section{Introduction}

\begin{figure}
    \includegraphics[]{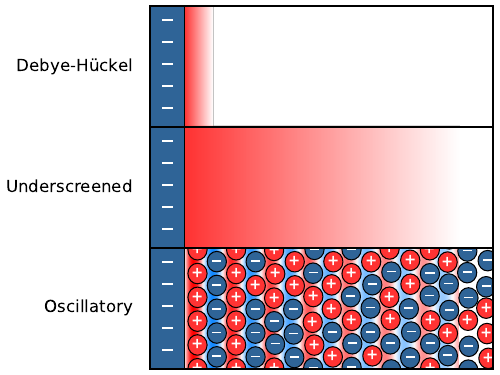}
    \caption{Differing theories of ion order at charged surfaces. The colored shade represents net charge distribution: red for positive and blue for negative. The electrostatic potential follows similar decay as well. Debye-Hückel theory treats ions as point charges without explicit ion-ion correlation and shows the shortest smooth decay. We can extend the simple decay to compensate for the overpopulation near the surface in Debye-Hückel theory and account for underscreening observed in experiments. Oscillatory decay arises when we include the effect of finite size of the ions. The discrete ions in the figure represent this nature that distinguishes this model from the other two.}
\end{figure}
Understanding nanoparticle interactions in concentrated electrolytes is a fundamental challenge in nanoscience, with direct significance for technologies ranging from electrochemical energy storage\cite{li_effect_2007,chen_ionogel_2018} to nanoparticle-enabled catalysis\cite{yan_ionogels_2024} and soft functional materials.\cite{fan_ionogels_2023} Ionic liquids, in particular, provide a highly tunable and strongly correlated environment, enabling the formation of complex materials such as colloidal ionogels.\cite{ueno_colloidal_2011, bideau_ionogels_2011} 
For decades, Derjaguin-Landau-Verwey-Overbeek (DLVO) theory has served as the standard framework for modeling colloidal interactions, combining van der Waals attraction and electrostatic repulsion into an effective pair potential.\cite{israelachvili_intermolecular_2011} This coarse-grained description is essential for bridging the large length and time scale disparity between nanoparticles and solvent molecules, which prohibits atomistic resolution.

DLVO theory relies critically on the so-called ``screening'' of nanoparticle surface charges by ions in the surrounding electrolyte. Classical models of the electric double layer (EDL), notably the Gouy-Chapman-Stern model, employ mean-field approximations to describe an exponential decay of electrostatic interactions over the Debye screening length.

However, extensive experimental evidence has shown that this picture breaks down in concentrated electrolytes and ionic liquids. Surface force measurements revealed that the characteristic length of the exponential decay of the electrostatic interaction, $\lambda_S$, becomes larger than the Debye length $\lambda_D$ at high ion concentration,\cite{gebbie_ionic_2013, smith_electrostatic_2016, gebbie_long_2017, lee_scaling_2017} leading to longer-ranged electrostatic surface interactions. These deviations are attributed to ion-ion correlations, the ion size and shape, and molecular ordering effects that are necessarily neglected by mean-field approximations.\cite{perkin_ionic_2012, rotenberg_underscreening_2018, elliott_known-unknowns_2024} Others have demonstrated that high ion concentrations can give rise not only to longer-ranged interactions, but also to oscillatory charge-density profiles associated with ion layering and self-assembly near charged surfaces.\cite{kirkwood_statistical_1936,kirkwood_statistical_1954, kjellander_dressed_1997, perkin_ionic_2012}

Despite significant theoretical and experimental progress, the microscopic origins and macroscopic consequences of these nonclassical EDL behaviors are not fully understood. Direct experimental probes, such as scanning probe techniques,\cite{schott_edl_2024} and X-ray-based methods\cite{wang_probing_2020} can resolve ion structuring near interfaces but are typically limited to only a few molecular layers and struggle to capture the fluctuating diffuse region. At the same time, atomistically resolving ion dynamics at nanoparticle surfaces remains prohibitively expensive for simulations of collective colloidal behavior. As a result, it is still unclear to what extent competing EDL theories influence emergent, mesoscale properties of colloidal materials.

This gap itself represents an opportunity: colloidal particles are experimentally-resolvable objects whose collective behavior can be characterized via optical microscopy. If the differences in ion ordering near nanoparticle surfaces lead to meaningful differences in colloidal interactions, then macroscopic structural and thermodynamic behavior can serve as an indirect indicator of interfacial electrostatics. 

In this work, we establish this connection by systematically investigating how competing models of electrostatic screening influence colloidal assembly in ionic liquids. Using molecular dynamics simulations of a model system with DLVO-type pair interactions parameterized by 1) classical Debye–Hückel decay, 2) longer-ranged monotonic decay (henceforth called ``underscreened"), or 3) oscillatory decay consistent with ion layering, we quantify differences in phase behavior, local structure, and mesoscale morphology. We show that changes in the form of electrostatic interactions lead to pronounced divergences in gel topology and thermodynamic signatures, even when local structural descriptors remain nearly indistinguishable. These results demonstrate that macroscopic colloidal behavior provides a sensitive probe of interfacial electrostatics and reveal that EDL models with different features lead to different predictions for nanoparticle assembly in concentrated electrolytes. We find that several other colloidal ionogels share similar differences in pair interactions arising from different electrostatic interaction models, implying the potential generalizability of this work.

\section{Results and Discussion}

\begin{figure*}
\begin{minipage}[t]{0.48\textwidth}
\centering
    \includegraphics[]{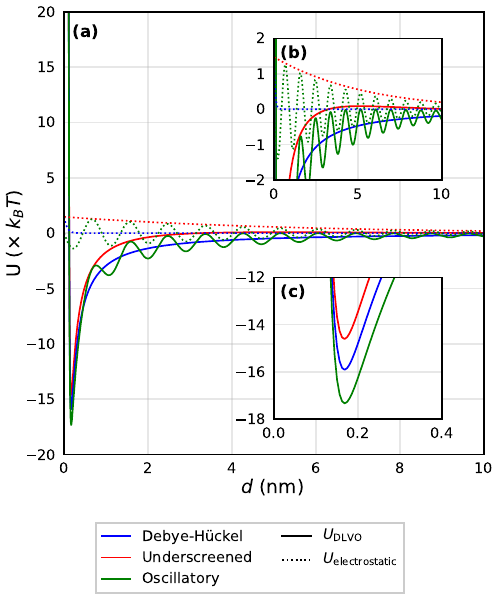}
    \caption{The DLVO pair potentials ($U_\text{DLVO}$, solid line) of our model colloidal system, resulting from the different electrostatic potentials ($U_\text{electrostatic}$, dashed line). The extension of electrostatic screening length from 0.05 nm in Debye-Hückel to 5 nm in Underscreened leads to the emergence of an energetic maximum around $d=5~\text{nm}$. Additional inclusion of oscillatory screening results in several energy maxima and minima. (b) and (c) show parts of (a) in detail. The equations and parameters used to derive the pair potentials are in Methods and Table \ref{tab:parameter}.}
    \label{fig:dlvo}
\end{minipage}
\hfill
\begin{minipage}[t]{0.48\textwidth}
\centering
\includegraphics[]{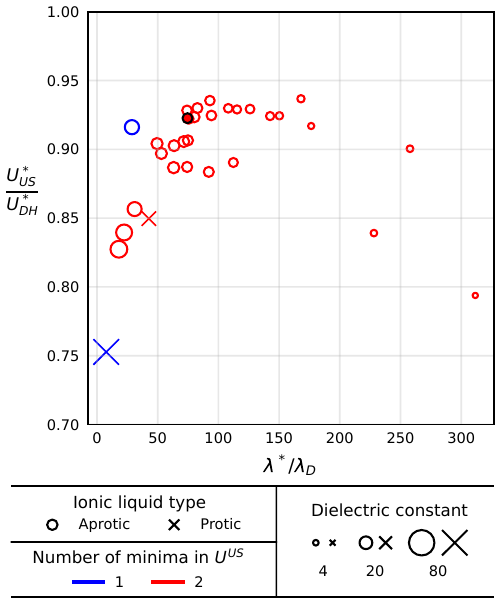}
\caption{The features of pair potential of 34 ionic liquids under the Debye-Hückel theory or the underscreened (US) model, assuming constant surface potential ($\phi_s=-40$~mV) condition. $\lambda_D$ is the Debye length and $\lambda^*$ is the decay length under the US model. The color represents the existence of a secondary minimum, hence a local maximum, of each pair potential. $U^*$ is the depth of the primary energy minimum, calculated by the difference between the primary minimum and either maximum (if energy maximum exists) or zero (if it does not exist). The point size represents the magnitude of dielectric constants. [EMIM][TFSI] is marked as green.}
\label{fig:il-diverse}
\end{minipage}

\end{figure*}

The following results are based on equilibrium NVT molecular dynamics simulations of three different electrostatic double layer theories, as incorporated into the DLVO potential (Fig.~\ref{fig:dlvo}) and performed over several simulation replicates. For our study, we assume bare silica nanoparticles ($r=100$ nm) submerged in an ionic liquid, 1-ethyl-3-methylimidazolium bis(trifluoromethylsulfonyl)imide ([EMIM][TFSI]), as a technologically-relevant\cite{chen_silica_2020} model system. The relevant properties are listed in Table.~\ref{tab:parameter}, which result in a screening length of roughly 0.05 nm (DH) and 5 nm (underscreened). Details on simulation setup and protocols can be found in the Methods section. Longer screening length under the underscreened model results in the emergence of a local energetic maximum at around $d=5~\text{nm}$ and the following secondary minimum (Fig. \ref{fig:dlvo}).

The trends discussed in the following sections have the potential to be generalized in colloidal systems with diverse ionic liquids, even though we performed simulations on a model system. As context, we collected bulk properties of 34 ionic liquids that had enough information to construct the DLVO-type pair potentials using the same equations we used for [EMIM][TFSI]. All the ionic liquids except for two---2-hydroxyethylammonium formate and tris(2-hydroxyethyl)methylammonium methylsulfate---demonstrate similar qualitative features as our model system, showing the existence of a maximum and a secondary minimum in their pair potentials (Fig. \ref{fig:il-diverse}). In Fig. \ref{fig:il-diverse}, $U^*_{US}/ U^*_{DH}$ represents the significance of the long-range electrostatic component in the pair potential; higher significance leads to a lower value, considering $U^*_{US}/ U^*_{DH} \to 1$ when the magnitude of electrostatic interaction is negligible compared to that of the van der Waals interaction. $\lambda^* / \lambda_D$ represents the extent of nonclassical deviation from the Debye-Hückel theory; larger deviation leads to a higher value. Most ionic liquids have similar or smaller $U^*_{US}/ U^*_{DH}$ compared to [EMIM][TFSI]. In addition, $\lambda^* / \lambda_D$ values are mostly similar to or larger than [EMIM][TFSI] and even many with smaller $\lambda^* / \lambda_D$ still show secondary minimum in their pair potentials. This implies that the different electrostatic interaction models can lead to different predictions in the phase and structural behavior of colloidal ionogels other than our model system.

\subsection{Energetics and phase behavior}
\begin{figure}
    \includegraphics[]{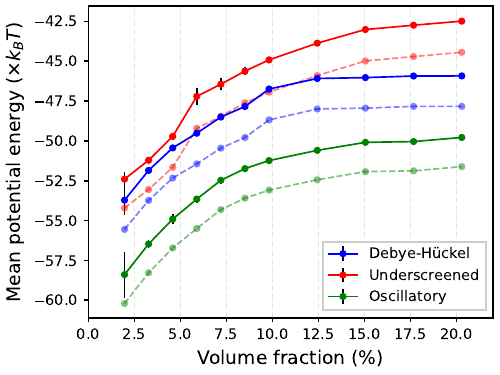}
    \caption{Potential energy per particle for different $U_\text{electrostatics}$ along the particle fraction. The solid lines are the potential energy values from molecular dynamics simulations. The dashed lines represent the multiplication of primary energy minimum in Fig.~\ref{fig:dlvo} and the normalized mean coordination number of the particles.}
    \label{fig:pe}
\end{figure}

To analyze how different electrostatic interaction theories impact collective colloidal phases, we investigated phase behavior through the analyses of the potential energies and the heat capacity. The potential energy per particle (Fig. \ref{fig:pe}, solid lines) shows an increase across particle fractions, indicating a decrease in the average coordination number as branching occurs in particulate gels. This is confirmed by multiplying the average coordination number and the value of the potential minimum (dashed lines), which shows that these two are the main factors affecting the potential energy.

\begin{figure}
    \includegraphics[]{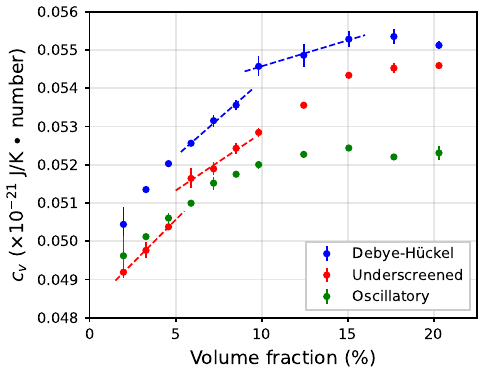}
    \caption{Intensive volumetric heat capacity ($c_V$) for different $U_\text{electrostatics}$ along the particle volume fraction. For Debye-Hückel and underscreened models, the discontinuity in $c_V$ is observed at the volume fraction between two separated lines. Each line is a least-square fit based on nine data points (three volume fraction values) it passes.}
    \label{fig:cv}
\end{figure}

The isochoric heat capacity ($c_V$, Fig. \ref{fig:cv}) provide a key understanding of phase boundaries and evolution. Since the simulations are conducted at a constant temperature, the observed discontinuity in $c_V$ directly indicate the discontinuity in the second derivative of the intensive Helmholtz free energy, implying the phase transition. In the Debye-Hückel case, this discontinuity occurs between $\phi=8.5\%$ and $9.8\%$, coinciding with the complete transition to a 3D-percolated network. In the longer-range decay (underscreening), we also observe a discontinuity in $c_V$, albeit at a much lower particle fraction ($\phi=4.6\% -5.9\%$). In contrast to the other models, the oscillatory model shows no distinct $c_V$ discontinuity.


After full percolation, we see a weaker signature of the phase transition between gel and glassy states, as signified by the leveling off of $c_V$ across all theories at $\phi=12.5\%-17.5\%$. Considering $c_V$ serves as a proxy for the system's available degrees of freedom, this still indicates that the systems develop into their more arrested, glassy structures where increase particle concentration restricts configurational freedom.

\subsection{Local configurations}

To quantify structural changes across different theories, we utilized two local descriptors: coordination number and Smooth Overlap of Atomic Positions (SOAP).\cite{bartok_representing_2013} Coordination number is a simple and intuitive way to describe the local configuration, but includes little information about the packing or distribution of neighboring particles. SOAP accounts for this nuance by placing a Gaussian distribution on each particle center and evaluating the resulting local particle density. SOAP has proven to be effective for distinguishing local environments for self-assembled materials,\cite{gardin_classifying_2022} making it well suited for characterizing local structure of our systems.




\begin{figure*}
    \includegraphics[]{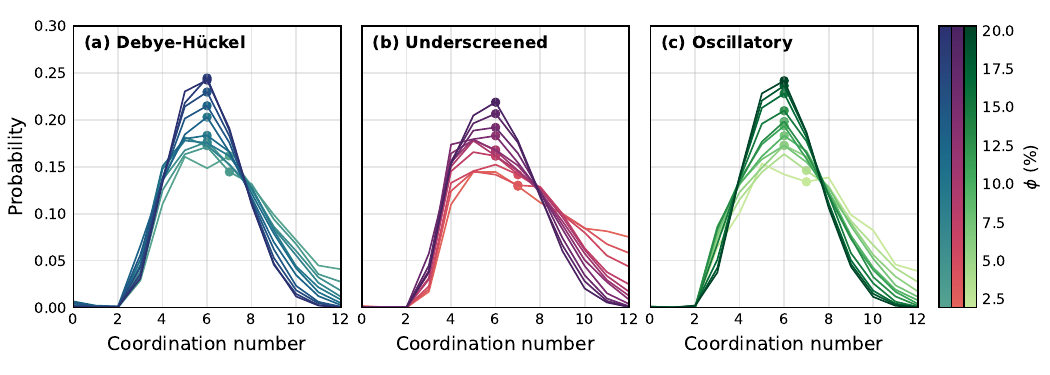}
    \caption{(a)-(c) Distribution of coordination numbers for different $U_\text{electrostatics}$ along the particle fraction. The darker colors represent higher particle fraction in each plot. The median coordination number for each particle fraction is presented as a dot with the same color on each line.}
    \label{fig:cn}
\end{figure*}
\begin{figure*}
    \includegraphics[]{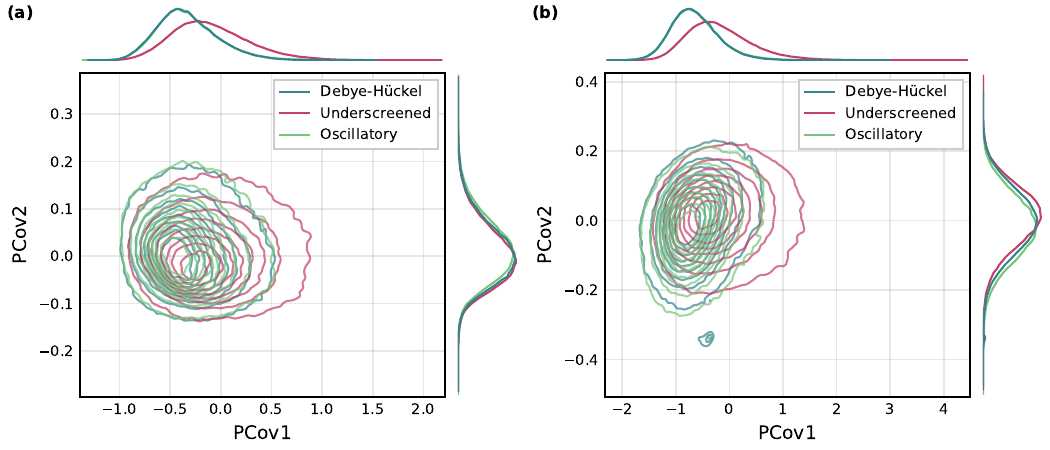}
    \caption{Distribution of local environments represented by SOAP and mapped by PCovC, based on (a) 1st ($r_{cut}=201\text{ nm}$) and (b) 2nd ($r_{cut}=401\text{ nm}$) neighbor shell analysis across different theories. PCovC provides a latent-space visualization of the data that best separates the distributions corresponding to each theory. In other words, this constitutes the ``viewing angle'' on the high-dimensional data that best separates the configurations resulting from each theory. The Debye-Hückel and the oscillatory potential show almost same distributions. The underscreened potential exhibits some differences along PCov1 axis, but mostly overlaps with the other two.}
    \label{fig:soap}
\end{figure*}


At low particle fractions, all three gels show flatter distributions of coordination numbers, corresponding to non-dense, variably-coordinated systems. As particle fraction increases, all three theories demonstrate a more well-defined peak near 6-8 neighbors, with a small difference in the average coordination number between the theories (with underscreening corresponding to approximately 0.1-0.4 more neighbors/particle). Part of this trend can be interpreted as an effect of the deeper primary minimum of the interaction potential (Fig.~\ref{fig:dlvo}), based on the same trend observed in a simulation study of a non-dense colloidal system under the Morse potential of different interaction strengths.\cite{mangal_small_2024} As the particle fraction increases, the coordination numbers of differing theories converge. To more exhaustively quantify local coordination, we employ SOAP descriptors, which provide a mathematical signature to the coordination environments, taking into account radial and angular components to the neighboring particles. Despite the sophistication of these descriptors and our employment of a mapping technology, PCovC, specifically intentioned to separate distributions in latent space\cite{jorgensen_interpretable_2026}, these local descriptors show the same qualitative similarities between theories. This is demonstrated in Fig. S1 (showing the PCA) and Fig.~\ref{fig:soap}, where the PCA shows full overlap of these statistical distributions, albeit with some small shifts for the underscreened model.


\subsection{Global topology}
Finally, we use the Euler Characteristic (EC), a feature vector that encodes the topology of the particle point cloud, as a metric to compare the resultant gel network morphologies. EC has been shown to be a powerful and information-dense description of several systems.\cite{smith_euler_2021, pranav_topology_2019, amezquita_measuring_2022} Importantly, it has been shown to be closely correlated with the mechanical and rheological properties of colloidal gels such as elastic and viscous moduli.\cite{smith_topological_2024} We randomly extract 10 snapshots after the equilibration period of each simulation and calculate their EC vectors. Three simulations with different random initial configurations were run for every (electrostatic model, $\phi$) pair, resulting in 30 EC vectors for each particle fraction and theory.

Notably, each snapshot corresponds to one EC vector, unlike the local descriptors that are assigned to a particle. The EC encodes the particle configuration as an alternating sum of $\beta_0$, the number of clusters, $\beta_1$, the number of two-dimensional holes, and $\beta_2$, the number of void spaces across a scan of neighborhood distances $r$ (called the ``filtration value'' in topological data analysis). For example, a small void space in the middle of a tetrahedron of touching particles is captured in $\beta_2$ at a short $r$, versus a large void resulting from the gel network is captured in $\beta_2$ at a longer $r$.

\begin{figure}
    \includegraphics[]{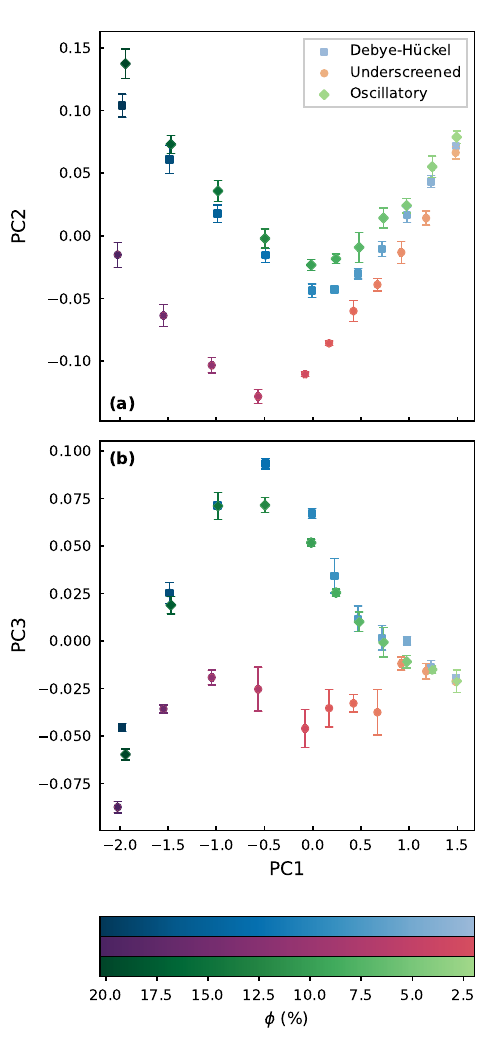}
    \caption{PCA of the Euler Characteristic vectors for different electrostatic theories across particle fractions ($\phi$). PC1 aligns with $\phi$, while PC2 and PC3 capture theory-specific variations perpendicular to $\phi$. All theories overlap at low $\phi$ but diverge at higher fractions. The underscreened potential differs significantly from the others. The Debye-Hückel and the oscillatory potentials remain relatively closer, still showing distinct differences especially at moderate $\phi$ ($5.9-9.8$\% in (a), $9.8-12.4$\% in (b)).}
    \label{fig:ec-pca}
\end{figure}

Contrary to the local descriptors, the EC vectors clearly demonstrate key similarities and divergences between the morphologies stabilized by each electrostatic theory, even when projected along principal components, even though PCA is not natively intended to separate these distributions. In Fig.~\ref{fig:ec-pca}, we show the first three principal components (PCs) of EC vectors for three electrostatic theories, where Debye-Hückel theory is in blue, underscreening theory is in red, and the oscillatory theory in yellow/green. The darker the color of the point, the higher the particle fraction and the error bars represent the standard deviation across data points for a given particle fraction and electrostatic theory. The first principal component clearly aligns with the particle fraction of each simulation, smoothly increasing from sol phases on the right to gel and glass phases on the left of Fig.~\ref{fig:ec-pca} (a). Our next two PCs, by definition, contain orthogonal information to the first PC and show clear divergences between theories – particularly between the Debye-Hückel and the oscillatory theories and the theory with the decay length of \citet{lee_scaling_2017}.


The differences in the second and third PCs among the theories become starker with a higher particle fraction, corresponding to the formation of longer-ranged structures. This trend disagrees with what the local descriptors show, where the differences decrease or remain similar along the particle fraction axis (Fig.~\ref{fig:cn}). This disagreement demonstrates that the structures of dense colloidal systems cannot be fully understood solely with these local descriptors that capture structural information up to the second neighbor shell. In other words, the local particle environment may not change, yet the global morphology, which is captured by EC, changes greatly between the different theories.

Surprisingly, the morphology derived from the oscillatory theory more closely aligns with that derived from Debye-Hückel theory. The same trend was observed in terms of structure factor $S(q)$ (Fig. S7). This is of note, as the oscillatory potential retains higher electrostatic repulsions and attractions at longer length scales, while the Debye-Hückel theory gives an electrostatic potential and force that become negligible quickly as we move from the particle surface. Still, their structural topologies show meaningful differences in PC2 at $\phi=5.9-9.8\%$ and PC3 at $\phi=9.8-12.4\%$. These $\phi$ regimes overlap with the lack of distinct discontinuity in $c_V$ under the oscillatory theory at $\phi=8.5-9.8\%$ (Fig. \ref{fig:cv}), implying that the DH and the oscillatory theory lead to different phase and structural transitions along the increase in $\phi$.

\begin{figure}
    \includegraphics[]{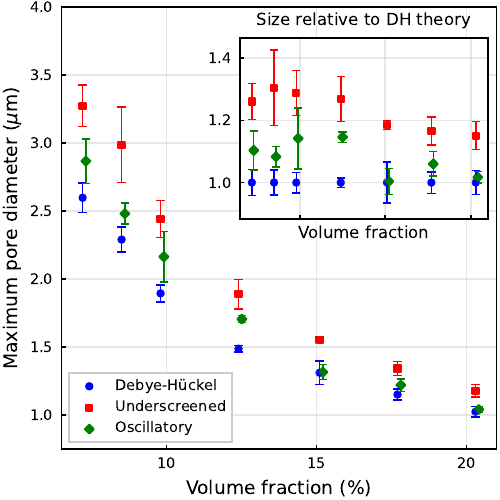}
    \caption{Maximum pore diameters of different electrostatic interaction models and particle volume fractions. The pore sizes were evaluated at 10 frames after equilibration timestep for each simulation, thus 30 distributions for each point plotted. Each point represents the mean and error bar represents the standard deviation of maximum pore diameters of corresponding distributions. The inset shows the relative sizes, the maximum pore diameters divided by the average value of those under the Debye-Hückel theory.}
    \label{fig:pore-max}
\end{figure}

These results, namely the difference in sol-gel phase transition and three-dimensional percolation, suggest that extended screening leads to more open network gels of significantly different topology than phases suggested by other theories. To confirm this, we identified the pore sizes of each system across particle fraction. Across all particle fractions, the extended electrostatic interaction model consistently yields larger maximum pore sizes than the Debye-Hückel theory (Fig. \ref{fig:pore-max}). The maximum pore sizes predicted by the oscillatory model fall between these two cases, approaching the Debye-Hückel results at high particle fractions. This trend corresponds to a systematic broadening of the pore size distribution toward larger volumes, rather than a few statistical outliers (Fig. S4). These findings codify that the extended range of electrostatic repulsion not only enables the formation of colloidal gels at lower particle fractions but also increases their openness relative to the shorter-ranged Debye-Hückel screening. This effect is presumably due to long-range repulsion between non-neighboring particles.

\section{Conclusions}
In this work, we have demonstrated that the specific form of the electrostatic interaction terms within DLVO-type potentials plays a decisive role in determining the phase behavior and structural morphology of colloidal systems in ionic liquids and concentrated electrolytes. By comparing three representative screening models: classical Debye Hückel theory, an underscreened single-decay model, and an oscillatory decay model, we have shown that differences in screening length and functional form lead to distinct phase boundaries, percolation behavior, and gel morphologies. 

We have demonstrated these differences through a variety of analysis and characterization techniques. First, we showed distinct differences in phase behavior across different electrostatic interaction models. The extended monotonic screening model demonstrated the discontinuity of $c_V$ at a lower particle fraction (around 5\%) compared to the Debye-Hückel theory (around 10\%), while the oscillatory screening model showed no distinct discontinuity. This implies different phase behavior for all the three models studied.
Local structural descriptors such as coordination number and SOAP vectors revealed only small differences between electrostatic models. Conversely, global analyses through persistent homology, captured through the Euler characteristic, demonstrated strong divergences in mesoscale topology that become pronounced at higher particle fractions. This is particularly interesting, as topological descriptors have been directly tied to the rheological properties of gel systems,\cite{smith_topological_2023} inferring that dynamical simulations of colloidal gels would reveal experimentally-measurable changes in gel rheology dependent on electrostatic models. Interestingly, oscillatory electrostatic interactions lead to gels topologically closer to those suggested by Debye-Hückel-type screening, yet, produce structural features distinct from the other two models. Finally, pore size analysis indicated that the increased screening length broadens the pore size distribution toward larger volumes. Hence, this work demonstrates that the formation of a highly open gel network can serve as a strong indicator of long-range interaction forces.

These findings also have important implications for computational simulations of colloidal materials. Implicit-solvent interaction models are widely used to enable simulations of large-scale systems at a tractable computational cost, particularly in the context of materials discovery where broad parameter spaces require high-throughput exploration. Our results suggest that the electrostatic assumptions embedded in these models strongly influence phase stability and structure, and the choce of screening length can artificially bias the results of simulations of implicit solvents representing charged electrolytes and solutions. In this regard, further studies on phase diagrams and gel formation mechanisms, spanning a wider range of temperature and particle fraction, would be valuable and provide more insights on the impact of long-range interactions onto colloidal behavior.\cite{ioannidou_crucial_2016}

More broadly, this sensitivity of collective colloidal properties to the underlying electrostatic model suggests that macroscopic observables, such as phase transitions, connectivity, and gel morphology, can serve as experimentally accessible fingerprints of interfacial ion ordering.
 Rather than relying on hard-to-obtain microscopic measurements of interfacial structure, future studies could combine experimental techniques such as scattering and confocal microscopy with simulations. They could provide a promising pathway for assessing and refining electric double-layer theories in complex ionic environments using experimentally accessible measurements.

\vspace{12pt}
\section{Methods}
\subsection{Implementation of the DLVO potential}
 To model interactions, we use functionalities and interaction models provided by LAMMPS.\cite{thompson_lammps_2022}
The van der Waals attraction term is

$$
    U_\text{vdw}(d) = - \frac{A}{6} \left[
 \frac{2 r^2}{4rd+d^2}
 + \frac{2 r^2}{\left(2r+d\right)^2}
   + \mathrm{ln}
     \left(
\frac{4rd+d^2}{\left(2r+d\right)^2}
 \right)
\right]
$$
%
where $r$ is the radius of the nanoparticle, $d$ is the closest separation distance between particle surfaces, and $A$ is the Hamaker constant, here assumed to be $1.583 \times 10^{-21}$ J for silica nanoparticles, derived from Lifshitz theory. \cite{israelachvili_13_2011} $$A=\frac{3}{4}k_BT \left( \frac{\varepsilon_S-\varepsilon_M}{\varepsilon_S+\varepsilon_M}\right)^2+\frac{3h\nu_e}{16\sqrt{2}}\frac{(n_S^2-n_M^2)^2}{(n_S^2+n_M^2)^{3/2}},$$  where $\varepsilon_M$ and $\varepsilon_S$ are the dielectric constant of the solvent and particle, respectively, $n$ is the respective refractive index, $k_B$ is the Boltzmann constant, and $T$ is the temperature of the system.
We introduce an additional steric term\cite{everaers_interaction_2003} to avoid particle overlap:

\begin{widetext}
\begin{equation}
\begin{split} U_\text{steric} =& \frac{A_{cc}}{37800}  \frac{\sigma^6}{(d+2r)}\left[ \frac{d^2-10dr+30r^2}
{d^7}               +\frac{d^2+18dr+86r^2}
{\left(d+4r\right)^7}  -2\frac{d^2+4dr-26r^2}
{\left(d+2r\right)^7}\right]
    \end{split}
\end{equation}
\end{widetext}
where $\sigma$ is the characteristic length scale for silica, here assumed to be 0.295 nm.\cite{anagnostopoulos_simplified_2021} For the electrostatic term, we use a tabulated potential derived from \citet{sader_accurate_1995} which yields the following:

\begin{widetext}
\begin{equation}
    \begin{split} 
  U_\text{electrostatic}(d) = {\varepsilon_M \varepsilon_0\left(\frac{kT}{e_0}\right)^2 Y^2(d)\left(\frac{r^2}{2r+d}\right)\ln{\left(1+e^{-d/\lambda_S}\right)}}  \end{split}
  \label{eq:3}
\end{equation}
\end{widetext}

where $$Y(d, \lambda_S) = 4e^{d/2\lambda_S} \tanh^{-1}{\left[e^{-d/2\lambda_S}\tanh{\left(\frac{y_s}{4}\right)}\right]}.$$ 

This potential term is largely governed by the ``screening length'', $\lambda_S$, the characteristic decay length of the electrostatic repulsion of the nanoparticles. For the $\lambda_S$, we evaluated the effect of different theories prevailing on these particles. Traditionally, the Debye-Hückel theory estimates the screening length as
$$\lambda_D = \left(\frac{\varepsilon_M \varepsilon_0  k_BT}{\sum_j{n_jq_j^2}}\right)^{1/2}$$
where $\{j\}$ includes all ionic species, $n_j$ is the number density of ion $j$, $q_j$ is the charge of ion $j$, and $\varepsilon_0$ is the vacuum permittivity.
This is the simplest representation of the screening length, assuming no ion-ion correlations, and thus assumes screening to directly follow the Debye length.
However, Debye-Hückel theory is a poor approximation when there exist many ion-ion correlations,\cite{elliott_known-unknowns_2024} as is common in highly ionic systems. As a result, $\lambda_D=0.06\text{ nm}$ for pure [EMIM][TFSI], which is smaller than the dimension of the constituent molecules.

The failure of Debye-Hückel theory is reflected as underscreening, a positive deviation of screening length from Debye length at high ion concentration. 
Here we incorporate this effect by setting an extended screening length $\lambda^*$ following an experimental scaling relationship\cite{lee_scaling_2017}, 
$$\frac{\lambda^*}{\lambda_D} \propto \left(\frac{a}{\lambda_D}\right)^{3/2}$$ 
 where $a$ is the size of the constituent ions. $a$ is calculated as $v^{1/3}$ where $v$ is half of the volume of an ion pair. If we set $\lambda^*/\lambda_D = \left({a}/\lambda_D\right)^{3/2}$, 
$\lambda^* = 4.56\text{ nm}$ for [EMIM][TFSI], which is on the same order of experimental value of $4.4-8.3\text{ nm}$.\cite{smith_electrostatic_2016,gebbie_long_2017, groves_wave_2024}
This value also strongly corresponds with the Bjerrum length, which represents the length scale of electrostatic correlations between two charges in a system.
The Bjerrum length of [EMIM][TFSI] calculated from the experimentally measured dielectric constant (Table \ref{tab:parameter}) is 4.6 nm. In this context, this relationship was used to estimate $\lambda^*$ and derive pair potential under the underscreened model for other ionic liquids in Fig. \ref{fig:il-diverse}.


Lastly, we consider the oscillatory nature of ion ordering at the nanoparticle surface suggested by \textit{Kjellander and Mitchell} \cite{kjellander_dressed_1997, kjellander_multiple_2020} on top of longer-range screening. In this approach, the effective force between nanoparticles is given as a function of a complex-conjugate inverse screening length $\kappa \pm \kappa^\prime i$:

$$f(d) = C\frac{\exp(-\kappa d)}{2r+d} \cos(\kappa^\prime d + \alpha)$$
where $C$ is a coefficient in units of energy, and $\alpha$ is a phase of charge oscillation. We construct $f(d)$ to act as an oscillatory analog to the potential defined by $\lambda^*$. Namely, we assume parameters that make the non-oscillatory part of $f(d)$ to follow $U_\text{electrostatic}(d)$, which corresponds to $\kappa^{-1}=\lambda^*$, $\kappa'=2\pi/1.5a$, $\alpha=\pi/2$.

This value of $\kappa'$ was determined by numerically solving the equations from \citet{kjellander_multiple_2020} to derive the complex decay length of  simple spherical ions of the same size:\cite{kjellander_multiple_2020}
$$\left[ \frac{\kappa'}{\kappa_D}\right]^2=\frac{e^{\kappa' a}}{1+a\kappa'}$$ where $\kappa_D=\lambda_D^{-1}$.
This force was then tabulated in place of $U_\text{electrostatic}$ for LAMMPS simulation. This serves to represent the resultant effects of introducing oscillations into slowly-decaying DLVO interactions.


\begin{table}
    \centering
    \begin{tabular}{|>{\centering\arraybackslash}p{0.15\linewidth}|>{\centering\arraybackslash}p{0.2\linewidth}|>{\centering\arraybackslash}p{0.5\linewidth}|}\hline
         Property&  Value& Definition\\\hhline{|=|=|=|}
         $\varepsilon_S$&  3.8& Dielectric constant of particle (amorphous silica)\cite{arpe_ullmann_1988}\\\hline
         $\varepsilon_M$&  12.3& Dielectric constant of medium ([EMIM][TFSI])\cite{krossing_why_2006} \\\hline
         $n_S$&  1.475& Refractive index of particle\cite{khlebtsov_determination_2008}\\ \hline
         $n_M$&  1.422& Refractive index of medium\cite{miran_beigi_investigation_2013}\\\hline
         $\nu_e$&  $3.00\times 10^{15}$& Main electronic absorption frequency (Hz)\cite{israelachvili_13_2011}\\\hline
         $\sigma$&  0.295& Dimension of constituent atoms of particle (nm)\cite{anagnostopoulos_simplified_2021}\\\hline
         $\psi_s$&  -0.04& Surface potential (V)\cite{ueno_colloidal_2008}\\ \hhline{|=|=|=|}
         $A$&  1.58& Hamaker constant ($10^{-21}$ J)\\\hline
         $\lambda_D$&  0.06& Debye length (nm)\\\hline
         $\lambda^*$&  4.6& Rescaled screening length (nm)\\\hline
         $\kappa$&  0.22& Real part of the complex conjugate inverse screening length (nm$^{-1}$)\\\hline
         $\kappa'$&  7.0& Imaginary part of the complex conjugate inverse screening length (nm$^{-1}$)\\\hline
    \end{tabular}
    \caption{Properties of the system materials, and the resulting screening lengths calculated for different models. Main electronic absorption frequency is assumed to be the same for both materials. The dimension of constituent atoms of silica is obtained by averaging the characteristic length of Lennard-Jones potential for Si and O in silica.}
    \label{tab:parameter}
\end{table}

\subsection{Simulation protocol}
We initialize systems ranging from 1,024-10,462 particles in random configurations for concentrations ranging between $2-20\%$ volume fraction of nanoparticles at room temperature (298.15K) employing a Nosé-Hoover thermostat. The well-dispersed random configurations were generated such that the surface distance between two particles is greater than 10 nm. The arrested systems, such as nonequilibrium colloidal gels, depend on the kinetic path chosen in the simulations such as an equilibration protocol. We aim to imitate the evolution of the system after ultrasonication, which is used to disperse colloidal particles in experimental studies,\cite{ueno_colloidal_2008, bertsch_interfacial_2019} by simulating these systems in the canonical (NVT) ensemble with periodic boundary conditions. To ensure no finite size effects, we queried system sizes $L=4-16~\mu$m to test for energy convergence, where energies converged at $L=6~\mu$m (Fig. S5). The cutoff radius for the pair potential was set to 260 nm, corresponding to $d=60~\text{nm}$, to include the long-range nature of the electrostatic component. The magnitude of $U_{DLVO}$ was less than $0.01k_BT$ for all models, and the differences in $U_{DLVO}$ between the models were negligible at the cutoff; in the order of $10^{-6}-10^{-5}k_BT$. 

All simulations were run via LAMMPS software, with the time of equilibration determined by potential energy convergence, employing the algorithms of \citet{chodera_simple_2016}. This resulted in simulations on the order of 20~$\mu$s$-$5~ms with a timestep size of 0.1 ns. We ran three separate replicas of each configuration at each volume fraction. Using the LAMMPS framework, we defined a unique trial as one that had a different random seed to initialize velocities and positions. This way, we ensured that a different particle configuration alone did not fundamentally change the final structure. With these completed trials, we extracted the thermodynamic statistics of the system, such as particle configuration and potential energy.

As a note, we do not implement Langevin dynamics; the random fluctuation force is significantly smaller than what is required to disrupt the steep pair potential well in Fig.~\ref{fig:dlvo}. Furthermore, the high viscosity of [EMIM][TFSI] ($32.42~\text{mPa}\cdot\text{s}$,\cite{liu_study_2021}) compared to other organic solvents would lead to prohibitively long equilibration times\cite{hijazi_fast_2018} and thus incorporating Langevin dynamics would incur significant computational overhead without much anticipated impact on equilibrium results. 

We conducted our simulation on two high-performance computing systems: Anvil\cite{anvil} at Purdue University through an allocation supported by NSF ACCESS,\cite{access} and a high-performance computing cluster at the Center for High Throughput Computing\cite{chtc} in the University of Wisconsin-Madison.

\subsection{Thermodynamic and structural definitions of phase phenomena}
The rigorous study of phase behavior in colloidal, disordered materials, is a challenge for molecular simulation and statistical theory, largely because of their dynamical behavior and amorphous morphology.  
There is considerable literature on characterizing phase phenomena in glassy or crystallizing systems,\cite{steinhardt_bond-orientational_1983, mickel_shortcomings_2013, spellings_machine_2018, martirossyan_local_2024, coslovich_dimensionality_2022, mau_stacking_1999, middleton_crystals_2001, fan_ideal_2026} which, while relevant to many problems, does not extend naturally to the problem of gelating systems studied via implicit solvent models. 

We investigate the phase behavior of our systems via response functions; this approach follows the classical Ehrenfest approach of determining phase boundaries based on divergences in free energy. 
Many thermodynamic derivatives, such as pressure and isothermal compressibility, are non-trivial or ill-founded in implicit solvent simulations.\cite{roux_implicit_1999} More straightforward is the isochoric heat capacity $C_V$, which is related to the second derivative of Helmholtz free energy and can be calculated from the fluctuations of internal energy of the system: $$C_V=-T \left( \frac{\partial^2 A}{\partial T^2} \right)_{V, N}=\frac{\left< E^2\right>-\left< E\right>^2}{k_BT^2}$$
For our simulations, we compute $\left< E^2 \right>$ and $\left< E \right>$ as moving averages with a window of 10,000 data points after the convergence of potential energies over simulation timeframes. We do this running average to avoid the problems of internal energy drift that can occur with slowly-equilibrating systems, as the fluctuations often converge well-before this timeframe (Fig. S6). $C_V$ was calculated as an average of the corresponding values calculated from all the windows. Intensive volumetric heat capacity $c_V$ is a number-averaged value of the extensive value: $c_V=C_V/N$.


\subsection{Local descriptors of state}
To quantify changes in structure across phase transitions, a classic approach is through structural or bond-order analysis; for non-crystallizing systems, however, these ``order parameters'' fall short of quantifying ``disordered'' structures.\cite{yang_determining_2021} 

For our local structure characterization, we employ two local descriptors: coordination number (CN) and Smooth Overlap of Atomic Positions (SOAP).\cite{bartok_representing_2013} 
While the former (CN) is a more standard approach, the latter (SOAP) assumes each particle represented as a density field and creates a structural signature by expanding over spherical harmonics and radial bases. Conceptually, this results in a vector which contains a sensitive description of the three-body correlation functions for any given particle.
The cutoff radius ($r_{cut}$) is set to 201 nm for both CN and SOAP to include the first neighbor shell of interactions. In order to analyze the structure of the second neighbor shell, the difference between SOAP vectors with $r_{cut}=401~\text{nm}$ and $r_{cut}=201~\text{nm}$ is evaluated.
The maximum number of radial basis function ($n$) and angular degree of spherical harmonics ($l$) are set to 8 and 6, respectively. These two parameters are resolution parameters of SOAP; that is, increasing $n$ and $l$ allows finer decomposition of particle density field, but it does not make a meaningful difference in the results.

High-dimensional SOAP descriptors were analyzed by reducing their dimensionality with Principal Component Analysis (PCA) and Principal Covariates Classification (PCovC).\cite{jorgensen_interpretable_2026} PCA maps high-dimensional data in a way that best preserves overall variance; PCovC is a hybrid mapping technique that additionally incorporates class information, herein defined as our choice of EDL theory. The importance ratio of preserving either variance or classes is determined by a mixing parameter, taken as 0.5 when both inputs (here the SOAP vectors) and labels (here the EDL theory) are properly scaled. In essence, PCovC shows how separable two distributions in a high-dimensional space are, and is here used to show that the distribution of nearest- and next-nearest-neighbors under different theories are near-identical in SOAP space. PCA and PCovC are computed via \texttt{scikit-learn}\cite{pedregosa_scikit-learn_2011} and \texttt{scikit-matter}\cite{ goscinski_scikit-matter_2025} Python packages, respectively. Due to the large amount of data, we use farthest point sampling\cite{cersonsky_improving_2021, goscinski_scikit-matter_2025} to reduce the data to train the model on while maintaining data diversity. \texttt{scikit-matter}\cite{ goscinski_scikit-matter_2025} was used to apply farthest point sampling to select 200 particles in each trajectory. The PCovC model is trained on the sub-selected data and then applied to the entire dataset of SOAP vectors.

\subsection{Global descriptors of state}

While local descriptors provide excellent insight into particle-level structures, mesoscale network structures are crucial to explain the properties of colloidal systems.\cite{nabizadeh_network_2024} In this sense, topological data analysis (TDA) provides a powerful way of analyzing morphology in various systems by quantifying the shape of complex data with computational topology.\cite{smith_topological_2021, smith_topological_2023, smith_topological_2024, dabaghian_topological_2012, townsend_representation_2020}

To apply TDA, we first construct geometric objects such as points, edges, and faces from a set of particle coordinates. A popular TDA method for this purpose is persistent homology, which defines the connectivity of the points by introducing a filtration value.\cite{otter_roadmap_2017} A simple example of filtration value is the radius of spheres centered at particle coordinates, where two data points are considered to be connected if the two corresponding spheres touch each other. The connectivity of the system evolves as we increase the filtration value, and persistent homology captures this evolution of topology by sweeping through a wide range of filtration values.

From this evolving network, we extract topological summaries. Among these, Euler characteristic (EC) is particularly useful because it is closely correlated to the mechanical properties of colloidal gels.\cite{smith_topological_2024} Furthermore, the EC serves as a robust descriptor because it is stable against noise, such as minor thermal fluctuations.\cite{otter_roadmap_2017} The EC is an alternating sum of Betti numbers $\beta_i$:
$$\text{EC}=\beta_0 - \beta_1 + \beta_2$$
where $\beta_0$, $\beta_1$, and $\beta_2$ represent the number of isolated clusters, two-dimensional holes, and three-dimensional voids, respectively. An EC value is defined for each filtration value, so each configuration yields a high-dimensional EC vector for a list of filtration values. Principal Component Analysis (PCA) is used to project these EC vectors into lower-dimensional space, allowing us to examine the topological changes of our colloidal systems.

In addition to EC, pore size distributions were determined by converting particle configurations into a voxelized representation using ASE \cite{larsen_ase_2017} and freud.\cite{freud2020} To capture complete pore distribution in periodic boundary condition, the particle coordinates were periodically replicated into a 2$\times$2$\times$2 supercell. Only the pores centered in the central 1$\times$1$\times$1 simulation box were considered. The structure was then discretized using cubic voxels of 20 nm side length. Each voxel was classified as ‘filled’ if its center is contained in one or more particles and as ‘void’ otherwise, yielding a binary field. The void space was then analyzed using image processing tools from SciPy \cite{virtanen_scipy_2020} and scikit-image.\cite{scikit-image} A Euclidean distance transform of SciPy was applied to calculate the distance to the nearest filled voxel from each void voxel, which represents the radius of the largest sphere that can be inscribed without overlapping the filled voxels. We then identified local maxima of the distance field which correspond to the pore centers using scikit-image. The resulting set of diameters was used to evaluate the pore size distribution.

The static structure factor $S(q)$ was computed from the molecular dynamics trajectories using the Debye scattering method implemented in the freud\cite{freud2020} package. Ten configurations were sampled from the equilibrated portion of each trajectory by selecting one frame from each of ten equal temporal intervals. Particle coordinates, normalized by the particle diameter $2r$, were used to calculate $S(q)$ in the range $(0.5\times \text{box length}~/~2r)^{-1} \le 2rq \le 1$.

\section{Acknowledgments}
Support for this research was primarily provided by the University of Wisconsin–Madison Office of the Vice Chancellor for Research via the Wisconsin Alumni Research Foundation, the National Science Foundation Graduate Research Fellowship Program under Grant No. DGE-1747503, and the National Science Foundation Division of Materials Research under Grant No. DMR-2309000. We also acknowledge support from the Michael and Virginia Conway Professorship. The funders played no role in study design, data collection, analysis and interpretation of data, or the writing of this manuscript.

Research was sponsored by the Army Research Laboratory and was accomplished under Grant Number W911NF-26-1-A040. The views and conclusions contained in this document are those of the authors and should not be interpreted as representing the official policies, either expressed or implied, of the Army Research Laboratory or the U.S. Government. The U.S. Government is authorized to reproduce and distribute reprints for Government purposes notwithstanding any copyright notation herein.

This work used Anvil at Purdue University through allocation CHM250049 from the Advanced Cyberinfrastructure Coordination Ecosystem: Services \& Support (ACCESS) program, which is supported by U.S. National Science Foundation grants \#2138259, \#2138286, \#2138307, \#2137603, and \#2138296.

\section{References}
\bibliography{references2}
\end{document}